\documentclass[conference]{IEEEtran}

\usepackage{listings}
\usepackage{hyperref}

\begin{document}

\title{MISO: An intermediate language to express\\parallel and dependable programs}

\author{\IEEEauthorblockN{Alcides Fonseca, Raul Barbosa}
\IEEEauthorblockA{CISUC, Department of Informatics Engineering\\
University of Coimbra\\
P-3030 290, Coimbra, Portugal\\
Email: \{amaf, rbarbosa\}@dei.uc.pt}
}
\maketitle

\begin{abstract}

One way to write fast programs is to explore the potential parallelism and take advantage of the high number of cores available in microprocessors. This can be achieved by manually specifying which code executes on which thread, by using compiler parallelization hints (such as OpenMP or Cilk), or by using a parallel programming language (such as X10, Chapel or \AE{}minium). Regardless of the approach, all of these programs are compiled to an intermediate lower-level language that is sequential, thus preventing the backend compiler from optimizing the program and observing its parallel nature.


This paper presents MISO, an intermediate language that expresses the parallel nature of programs and that can be targeted by front-end compilers. The language defines `cells', which are composed by a state and a transition function from one state to the next. This language can express both sequential and parallel programs, and provides information for a backend-compiler to generate efficient parallel programs. Moreover, MISO can be used to automatically add redundancy to a program, by replicating the state or by taking advantage of different processor cores, in order to provide fault tolerance for programs running on unreliable hardware.

\end{abstract}

\IEEEpeerreviewmaketitle

\section{Introduction}

In recent years, processor architectures have evolved towards processors with multiple CPU cores, rather than increasing the operating frequency. In order to improve the performance of programs, developers have to design (or redesign) programs to execute code in parallel. Writing parallel code can be done through the usage of libraries, such as Pthreads, or with special parallel programming-languages. In any of the cases, the parallel code is translated to a sequential intermediate language. Regardless of being compiled or interpreted, the intermediate language can be used to perform optimizations at the lower-level code. These optimizations are unaware of the parallel nature of the program and cannot take into account.

Intermediate languages handle a program as a list of operations, organized in blocks expected to be executed sequentially. Parallel programs are also considered to be sequential, and the parallel nature of a program is hidden behind system calls for which the compiler does not know the semantics. This is true for LLVM, B3 and Swift intermediate languages, which are compiled, as well for virtual machine bytecode languages, such as Java bytecode, .NET Intermediate Language, Python bytecode and others.

Since these intermediate languages are shared between several high-level front-end programming languages, they are the focus of optimization. Parallelization is one type of optimization that can result in very significant speedups for large programs. Polly~\cite{grosser2011polly} is a tool that performs automatic parallelization of loops at the LLVM-IR level. Universal Translation Library~\cite{drozdov2014program} is another approach that takes LLVM-IR, translates to its own representation of the parallel program and then generates the resulting LLVM-IR. Both these approaches use internal representations that are not intended to be a target of compilation. 

SPIRE~\cite{khaldi2012spire} has tackled this issue by extending LLVM with \textit{detach}, \textit{attach} and \textit{barrier} constructs that allow asynchronous execution. This simple extension makes LLVM able to represent parallel languages such as OpenCL, Cilk, OpenMP, X10 and Chapel, while maintaining some lower-level semantics. Despite this effort, there are some other parallel semantics not being represented, such as parallel for-loops and task dependencies, among others. The limited semantics of SPIRE prevent some optimizations that could have been done otherwise. 

In this paper, we present MISO, an intermediate language for representation of parallel programs, supporting both Single Instruction Multiple Data (SIMD) and Multiple Instructions Multiple Data (MIMD). MISO can be used for performing optimization at the parallelization level, as well as for improving the dependability of programs running on unreliable hardware.

\section{A MISO Primer}

MISO stands for Multiple-Input Single-Output, which describes the semantics of each MISO block. MISO is organized in `cells', of which there can be several instances. A cell has two components: the state definition and the state transition function.

Listing~\ref{code:blend} shows the example code of a program that progressively blends one image into another. The ImageBlend cell is defined as having three memory slots: r, g and b, all integers (the state). That cell also has a transition function that performs a weighted sum of the value of each color in the previous state, and the same color from the second image in the previous state as well. The saved values are stored in a new state that will be feed into the next transition. In the bottom of the program, we can see the instantiation of several cells, one for each pixel of the two images. The source code for StaticImage is omitted, as it is similar to ImageBlend with an empty transition function.

\begin{lstlisting}[language=Java, caption={An example of a MISO program}, label={code:blend}]

cell ImageBlend {
  var r:Int = 0;
  var g:Int = 0;
  var b:Int = 0;
	
  transition {
    r = .99 * r + .01 * image2(this.pos).r;
    g = .99 * g + .01 * image2(this.pos).g;
    b = .99 * b + .01 * image2(this.pos).b;
  }
}

image1 = new ImageBlend(300*200)
image2 = new StaticImage(300*200)
\end{lstlisting}

One aspect of MISO is that loading input and output data can be performed by the runtime. In this case, the program could load the two images from file and could output all intermediate states to screen, resulting in a video animation transition from one image to another. 

The state transition moves each cell from one state to the next. In order to achieve this safely, there can be only writes to the current state, or local variables. Reads can be performed from the previous state of either the current cell or any other cell. This semantic restriction allows for further optimizations.

\section{Parallelization of MISO}

One of the main challenges for automatic parallelization is the extraction of data-flow information, in order to understand what computations can influence further operations, in order to establish dependencies and guarantee that the parallel program has the same semantics of the sequential version. MISO describes those dependencies explicitly in the transition function. Cells that have dependencies can be synchronously executed, while cells without any direct or indirect dependency can be executed in parallel, removing the need for a global barrier per transition step.

Furthermore, both task and data parallelism can be described in MISO. Task parallelism (Multiple Instructions, Multiple Data) can be represented using different types of cell and Data parallelism (Single Instruction, Multiple Data) can be expressed by having several instances of the same cell. Data parallelism in MISO can be directly translated to GPU languages, such as CUDA or OpenCL.

\section{Dependability of MISO}

Integrated circuits are continually evolving toward higher density and greater number of transistors, as well as smaller gates and lower operating voltages. Although this trend vastly increases the performance of microprocessors, it also increases the rate at which soft errors occur. Soft errors are incorrect states in the hardware, caused by transient events such as particle strikes, that must be corrected in order for programs to produce correct results. We introduce redundancy in MISO in order to tolerate hardware errors. 

In MISO, all state transitions are isolated from one another. Since the state of one cell is only written by one state-transition function, replication of operations can be achieved through the same means as parallelization. Given that each cell has its own memory, the memory contents may be duplicated and the state-transition function can operate on both replicas. This replication can be applied in parallel in different CPU cores or processors. In NUMA architectures, the state duplication can be achieved even in different memories. Identifying a soft error can be done by comparing the two new states. If there is a mismatch, a third equal transition should be executed to decide between the two possible outcomes. By identifying MISO cells that are frequently erroneous, it is possible to detect permanent failures in hardware requiring maintenance.

Replication is managed by the runtime environment, and therefore the same MISO program may be executed with different levels of redundancy. In some circumstances a given piece of code may be crucial for a specific application, thereby requiring full duplication and comparison of MISO cells, while the same piece of code may play a secondary role in a non-critical application and require no replication. Selective replication of key cells may also be applied by the runtime, in order to balance the fault tolerance and the overhead.


\section{Conclusion}

This paper presented MISO, an intermediate language that can represent programs with both task and data parallelism implicit in its semantics. By isolating state and state transitions, compiler tools can reason about the program in order to perform optimizations in terms of parallelization. Furthermore, the state isolation of MISO cells allows for replication in different hardware processors and memories, in order to detect and recover from transient hardware faults.

The MISO toolset is currently under development, and a prototype is available at \url{https://github.com/alcides/miso-dsl}, featuring a sequential and parallel runtime execution system. The toolset is open source and open to improvements and suggestions from the community.

\section*{Acknowledgment}

The first author was supported by the Portuguese National Foundation for Science and Technology (FCT) through a Doctoral Grant (SFRH/BD/84448/2012).

\bibliographystyle{IEEEtran}
\bibliography{references}
\end{document}